\documentstyle[12pt]{article}

\begin{document}

\title{BLACK HOLES AND  QUANTUM MECHANICS}

\begin{titlepage}
\begin{flushright}
 {\tt hep-th/9902007}\\
January 31 1999
\end{flushright}
\vfill
\begin{center}
{\Large \bf {BLACK HOLES AND \\

\

 QUANTUM MECHANICS}}\vskip 0.3cm

\vskip 10.mm
 Renata
Kallosh  \\
\vskip 0.8 cm
 Physics Department, \\ Stanford
University, Stanford, CA 94305-4060, USA\\[2mm]
\end{center}
\vfill
\centerline{\bf Abstract}

\

The motion of a particle near the RN black hole horizon  is described by
conformal mechanics. Models of this type have no ground state with  
vanishing
energy. This  problem was resolved in past by a redefinition of the
Hamiltonian which  breaks translational time invariance  but gives  a
normalizable ground state. We show that this change of the  
Hamiltonian is a
quantum mechanical equivalent of the change of coordinates near the  
black hole
horizon removing the  singularity. The new Hamiltonian of quantum  
mechanics is
identified as an operator of a  rotation between 2 time-like  
coordinates of the
adS hypersurface which translates
global time.  Therefore conformal quantum mechanics may eventually   
help to
resolve the  puzzles of the classical black hole  physics.

\vfill
 \hrule width 5.cm
\vskip 2.mm

To appear in proceedings of the 22nd Johns Hopkins Workshop,  
Gothenburg, 1998.
\end{titlepage}

 1. Recently a surprising connection \cite{CDKKTV} between black holes and
conformal mechanics of De Alfaro, Fubini and Furlan \cite{DFF} (DFF)  and
superconformal \cite{AP} mechanics  have been established\footnote{Various
aspects of black holes and $ADS_2/CFT_1$ duality were studied in  
\cite{Mald}.}.
The dynamics of a (super)particle near the horizon of an extreme
Reissner-Nordstr\"om black hole was shown to be governed by
an action that reduces to a (super)conformal mechanics model
in the limit of a large black hole charge.  The Hamiltonian, and the  
rest of
the generators of the conformal group were found in \cite{CDKKTV} to  
be equal
to
\begin{equation}\label{dffham}
H= {p^2\over 2f} + {g\over 2x^2f}\, , \qquad K= f {x^2\over 2} \ ,  
\qquad D=
{xp+px\over 4} \ .
\end{equation}
 Here the function $f(x,p)$ in the limit of the large black hole  
charge $Q$
tends to 1
and  we find a complete agreement with the conformal model \cite{DFF}
in this limit.
Such models upon quantization give the commutators of the conformal  
algebra
\begin{equation}
[H, D] = iH \ , \qquad [K, D] = -iK \ , \qquad [H, K] = 2iD \ .
\end{equation}
This quantum mechanical model was shown in \cite{DFF} to a have a  
continuous
spectrum of energy
eigenstates with energy eigenvalue $E>0$, but there is no ground state
at $E=0$.

In ref. \cite{CDKKTV} a black hole interpretation of this model was  
suggested.
The classical analog
of an eigenstate of $H$ is an orbit of a timelike Killing vector  
field $k$,
equal to $\partial/\partial t$ in the region outside the horizon, and the
energy
is then the  value of $k^2$. The absence of a ground state of $H$ at  
$E=0$ can
now be interpreted as due to the fact that the orbit of $k$ with  
$k^2=0$ is a
null geodesic generator of the event horizon, which is not covered by the
static
coordinates adapted to $\partial_t$.

The procedure used by DFF to
cure the problem of the absence of a ground state was to choose a  
different
combination of conserved charges as
the Hamiltonian.  From the perspective of the quantum mechanics this  
shift of
the Hamiltonian is unusual and motivated by the fact that when the  
theory does
not have a ground state, an additional prescription may be used such  
that the
new theory has a well defined vacuum. It has been realized that such  
additional
prescription leads to the breakdown of  time translational   
invariance in favor
of maintaining a well-defined vacuum and an anti de Sitter group.

{}From the perspective of the particle motion near the black hole  
horizon, the
procedure adopted in \cite{DFF} is extremely well motivated. It was  
stated in
\cite{CDKKTV} that it  corresponds to a different choice of time,  
one for which
the worldlines of static particles pass through the black hole  
horizon instead
of remaining in the exterior spacetime.

\

2. We will discuss here  the connection between the black hole horizon and
absence of a normalizable  lowest eigenstate of a quantum mechanical
Hamiltonian in more detail.
By studying the conformal quantum mechanics   DFF reached the  
conclusion that
the Hamiltonian of the conformal quantum mechanics is not acceptable for
solving the problem of motion.
They suggested to solve the problem of motion
to use a compact operator $L_0$, which is not a Hamiltonian but a  
combination
of the Hamiltonian $H$ and conformal operator $K$, \begin{equation}
L_0= {1\over 2} \left({1\over a}K + aH \right) = {a\over 4 } \left(  
{x^2\over
a^2} +  p^2 + {g\over x^2} \right) \ .
\end{equation}
This compact operator was shown to exhibit the pleasant features of having
 a discrete spectrum and normalizable eigenstates. Here $a$ is a  
constant (with
dimension of length) whose introduction breaks scale invariance. The  
action is
scale invariant but the ground state is not, so that one may qualify  
this as a
spontaneous breaking of scale invariance.

It has been conjectured in \cite{DFF} that the constant $a$ may have a
fundamental meaning since it  provides a sort of infra-red cut-off  
responsible
for the good behaviour at large distances and the discrete spectrum.  
In fact it
was found in \cite{DFF} that from three independent operators  
$H,K,D$ only one
combination can be made which corresponds to a generator of a  
compact rotation.
The other two operators,
\begin{equation}
L_{\pm} = {1\over 2} \left ({1\over a}K - aH\right)  \pm iD
\end{equation}
correspond to hyperbolic non-compact transformations.  The algebra  
of conformal
generators   $H,K,D$ generators was replaced by the algebra of their  
linear
combination,
\begin{equation}\label{comut}
[L_0, L_{\pm}] = \pm L_{\pm} \ , \qquad [L_+, L_-] = -2 L_0 \ .
\end{equation}
The eigenvalues and eigenstates of the operator $L_0$ have been found and
$L_{\pm}$ were treated as raising and lowering operators.
The eigenvalues of $L_0$ are given by a discrete series $r_n = r_0  
+n$ where
$r_0= {1\over 2} (1+ \sqrt{g+1/4})$.

The superconformal mechanics \cite{AP} faced the analogous problems of the
absence of the well defined ground state and therefore in  
supersymmetric case
the same approach as in non-supersymmetric case was
developed, i.e. the shift of the Hamiltonian was performed. Moreover  
in the
fermionic part it was necessary to find the appropriate combinations of
supersymmetry and special conformal supersymmetry to provide the  
existence of
the ground state with the vanishing energy.

The worrisome part of this procedure invented 22 years ago  was the  
breakdown
of time translational invariance which in a  quantum mechanical  
system  in the
flat space is not clearly justified (apart from the fact that it  
helps to find
the vacuum and the physical states of the theory).

DFF procedure of an unusual but successful treatment of conformal quantum
mechanics
comes out in a new light from the ADS/CFT perspective.
The main difference with the old picture is that we start with the  
curved space
of a black hole where it is not known how to describe quantum  
mechanically the
motion of a particle near the horizon and  in the black hole  
interior. However,
the fact that the time translational symmetry is broken in the coordinate
system where the horizon looks like a singularity is not surprising  
at all!
Quite opposite, there is no translational symmetry in this situation  
in curved
space since near the horizon the time coordinate is ill defined, as  
will be
shown below in detail.

Note that in  classical general relativity as different from quantum  
mechanics
it is a standard procedure   to demonstrate that the horizon is not  
a  true
singularity by changing the coordinate system.

\

3. Let us  recall here the  following \cite{DGT}
  adS-type explanation as to why the extreme black hole horizon is  
not a true
singularity. The metric of the extreme RN metric in isotropic  
coordinates is
\begin{equation}
ds^2 = -\left(1+ {|Q|L_P\over \rho}\right)^{-2}dt^2 + \left(1+  
{|Q|L_P\over
\rho}\right)^2\big[d\rho^2 + \rho^2 d\Omega^2\big]\,,
\end{equation}
where $Q$ is the black hole charge, $L_P$ is the Plank length
,$d\Omega^2=d\theta^2 + \sin^2\theta \, d\varphi^2$ is the
$SO(3)$-invariant
metric on $S^2$, and $M= |Q|/L_P$ is the black
hole mass. The near-horizon (or large charge) geometry is
therefore
\begin{equation}
ds^2 = -\left({\rho\over QL_P}\right)^2 dt^2 + \left({QL_P\over  
\rho}\right)^2
d\rho^2
+(QL_P)^2d\Omega^2\,,
\end{equation}
which is the Bertotti-Robinson (BR) metric. It can be
characterized as the $SO(1,2)\times SO(3)$ invariant  
conformally-flat metric on
$adS_2\times S^2$. The parameter $QL_P$ may now be interpreted as  
the $S^2$
radius
(which is also equal to the radius of curvature of the $adS_2$ factor).
In horospherical coordinates $(t,\phi=\rho/QL_P)$ for $adS_2$, the
4-metric  of the BR solution of Maxwell-Einstein theory is
\begin{eqnarray}
ds^2 &=& -\phi^2 dt^2 + {(QL_P)^2\over \phi^2}\, d\phi^2 + (QL_P)^2  
d\Omega^2
\,.
\label{BR}\end{eqnarray}
The metric is singular at $\phi=0$, but this is just a coordinate
singularity and $\phi=0$ is actually a non-singular degenerate  
Killing horizon
of
the timelike Killing vector field $\partial/\partial t$. To see this  
one may
define an $adS_2$ space as {\it a surface in a flat space with one  
more time
variable and the $SO(1,2)$ metric}

\begin{equation}
  X^{\hat m}\eta_{\hat m \hat n} X^{\hat n}  + a^2 = - (X^0)^2 -X^+  
X^-+ a^2 =0
\ .
\label{adS}\end{equation}
One may introduce the hypersurface coordinates $(\phi, t)$ by choosing

\begin{equation}
X^0 = \phi t \ , \qquad X^- = a \; \phi \ , \qquad X^+ = {a^2 -t^2  
\phi \over a
\;  \phi} \ .
\end{equation}
The metric induced on the surface is the BR metric (\ref{BR}) where    the
radius of the adS space $a$ is related to the black hole mass and  
charge as
follows:
\begin{equation}
a= |Q_{BH}| L_P = M_{BH} L_P^2 \ .
\end{equation}
The  horospherical coordinates do not cover all of $adS_2$ but only  
a half of
it, they cover ${adS_2\over J}$ where $J$ is the antipodal map. At  
$\phi =0$
the metric (\ref{BR}) has a singularity related to the black hole  
horizon and
one has to choose either $\phi>0$ or $\phi<0$. The coordinate $X^-=  
a \; \phi $
 can be positive as well as  negative, but if $\phi$ is restricted to be
positive, this also means that $X^-$ has to be positive, i.e. take  
half of the
values which it is allowed when considered as defining a surface  
(\ref{adS}).

The coordinates $X^0, X^+, X^-$ evaluated on the hypersurface are smooth,
therefore $\phi$ remains a good coordinate at the horizon $\phi=0$,  
while the
time coordinate $t$ becomes ill defined. This means that the  
original black
hole metric near the horizon has to be described using different  
coordinates.
Let us first introduce a complex variable $T$ which is a complex  
combination of
two time-like coordinates,   \begin{equation}
T= T_1+ iT_2 \ , \qquad T_1\equiv {X^+ + X^- \over 2} \ , \qquad  
T_2\equiv X^0\
{}.
\end{equation}
A space-like coordinate is
\begin{equation}
r\equiv {X^+ - X^- \over 2} \ .
\end{equation}
In terms of coordinates (\ref{global}) the surface is
\begin{equation}
-(T_1)^2 - (T_2)^2  +r^2 +a^2  = T \,T^* +r^2 +a^2 =0 \ .
\end{equation}
One may use a different solution to the surface constraint,
\begin{eqnarray}
T=  e^{i \tau\over a}  \left (a^2+ r^2\right)^{1/2}
\label{global}
\end{eqnarray}
In these coordinates the metric is
\begin{eqnarray}
ds^2 &=& -\left ({a^2 + r^2\over a^2} \right) d\tau^2 +\left ({a^2  
\over a^2 +
r^2} \right)  dr^2  + a^2 d\Omega^2
\,.
\label{BR2}\end{eqnarray}

\

4. The $SO(1,2)$ symmetry is linearly realized on the coordinates of the
surface $X^{\hat m} \eta_{\hat m \hat n} X^{\hat n} +a^2=0$ as
$
 \delta X^{\hat m} =   \Lambda^{\hat m} {}_{\hat n} X^{\hat n}
$.
No translations are allowed, only rotations. The generators of this  
symmetry
$ J^{\hat m \hat n}= i X^{[\hat m} \partial ^{\hat n]}$  form the  
$SO(1,2)$
algebra:
\begin{equation}
[ J_{\hat m \hat n} , J_{\hat p \hat q}]= i ( \eta _{\hat m[ \hat p}  
J_{\hat q]
\hat n}-\eta _{\hat n[ \hat p} J_{\hat q] \hat m}) \ .
\label{lor}\end{equation}
The $SO(1,2)$ symmetry consists of a compact rotation between two  
time-like
coordinates,   $T_1$ and $T_2$   and of the two boosts between the  
space-like
coordinates $r$ and two times. The rotation between two times can be  
easily
exponentiated by considering it as a $U(1)$ transformation
\begin{equation}
T' = e^{i\alpha} T
\end{equation}
This transformation is generated by the shift of the global time  
$\tau' = \tau
+a \alpha $ so that
\begin{equation}
T' = e^{ i\tau'\over a}  \left (a^2+ r^2\right)^{1/2}  = T' =  
e^{i\alpha} T\ .
\end{equation}
Two boosts
\begin{equation}
\delta T = \Lambda r  \qquad \delta r ={1\over 2}(  \Lambda T^* +  
\Lambda^* T)
\end{equation}
generate a $U(1)$ transformation. The time translation generator is  
$L_0$ and
the complex boost  is generated by $L_{\pm}$.

In terms of $SO(1,2)$ transformations (\ref{lor}) the generators of the
quantum mechanics with the well defined ground state are given by
\begin{equation}
 J_{T_1 T_2} =L_0 \qquad J_{T_1 r }\pm i J_{T_2 r } = L_{\pm}
\end{equation}
Thus we described an  interpretation of the quantum mechanical  
operators $L_0,
L_{\pm}$ starting with black holes and changing coordinates near the  
horizon to
avoid a singularity.

The new Hamiltonian $L_0$ of the conformal quantum mechanics does generate
translation of the time $\tau$ in the coordinate system describing  
the extreme
black hole near the horizon, where the time $t$ of the horospherical  
coordinate
system is ill defined. Somehow the conformal quantum mechanics found  
the  way
to a correct description of the physics
by breaking $t$-time translation symmetry which is now understood in the
context of  the black hole horizon.

The black hole interpretation suggested in \cite{CDKKTV} of the unusual
features of a quantum mechanical system  is clear. One may want   to  
use this
correspondence in the opposite direction, to resolve the  problems  
of black
holes of the classical general relativity by some kind of  
generalized quantum
mechanics which may have somewhat unusual features as in the example  
of the
conformal DFF model, described above. A quantum mechanics of such  
nature may
tell us about the black hole interior.

\

I had clarifying  discussions of the issues of black holes and quantum
mechanics with G. Gibbons, A. Strominger, L. Susskind  and P.  
Townsend. I am
grateful to the organizers of the conference "Novelties in String  
Theories"  in
Gothenburg for the stimulating environment.  This work is supported  
by the NSF
grant PHY-9870115.

\end{document}